# Experimental behaviour of a three-stage metal hydride hydrogen compressor

A.R. Galvis E.[1], F. Leardini[1], J.R. Ares[1], F. Cuevas[2], J.F. Fernandez[1]

[1] MIRE-Group, Dpto. Física de Materiales, Facultad de Ciencias, Universidad Autónoma de Madrid, 28049 Madrid, Spain.
[2] Univ Paris Est Creteil, CNRS, ICMPE, UMR7182, F-94320, Thiais, France.

E-mail: josefrancisco.fernandez@uam.es

## Abstract

A three-stage metal hydride hydrogen compressor (MHHC) system based in $AB_2$-type alloys has been set-up. Every stage can be considered as a Sieverts-type apparatus. The MHHC system can work in the pressure and temperature ranges comprised from vacuum to 250 bar and from RT to 200ºC, respectively. An efficient thermal management system was set up for the operational ranges of temperature designed. It dumps temperature shifts due to hydrogen expansion during stage coupling and hydrogen absorption/desorption in the alloys. Each reactor consists of a single and thin stainless-steel tube to maximize heat transfer. They are filled with similar amount of $AB_2$ alloy. The MHHC system was able to produce a compression ratio (CR) as high as of 84.7 for inlet and outlet hydrogen pressures of 1.44 and 122 bar for a temperature span of 23 to 120°C.

## 1. Introduction

The pursuit of sustainable and renewable fuels has increased the research on novel technologies. Between several choices, hydrogen is an interesting energy vector due to its high energy content, diversified sources of supply and zero greenhouse gas emissions when produced using primary renewable energies [1], [2]. On this regard, hydrogen compressors provide an excellent approach to integrate the production and storage with the distribution of hydrogen as an energy carrier for different applications [3 - 8]. Specifically, thermally-driven Metal Hydride Hydrogen Compressors (MHHC) have significant advantages in comparison to other compression technologies and therefore have concentrated much attention in recent years [3], [4].

Several prototypes and experimental approaches have been developed to enhance the performance of this technology. Laurencelle et al. [9], designed a three-stage $AB_5$ MHHC prototype for a hydrogen production facility using a low-pressure alkaline electrolyzer. The $AB_5$ intermetallic compounds selected for the system were $LaNi_{4.8}Sn_{0.2}$, $LaNi_5$ and $MmNi_{4.7}Al_{0.3}$ for the first, second and third stage respectively. All stages were thermally cycled between 20 and 80 °C. The system supplied a flow rate of about 20 L (NTP) of hydrogen per hour. Vanhanen et al. [10] studied the feasibility of combined MHHC and heat pump devices through the characterization of various alloys. For both devices, $AB_2$-type Hydralloy C2 ($Ti_{0.98}Zr_{0.02}Mn_{1.46}V_{0.41}Cr_{0.05}Fe_{0.08}$) and C0 ($TiMn_{1.5}V_{0.45}Fe_{0.1}$) were used for the first and second stage, respectively. The hydrogen pressure increased from 12-18 bar up to 85-110 bar by utilizing a narrow temperature range (20-60ºC). Moreover, when adding a third stage with $AB_2$ alloy ($TiCrMn_{0.55}Fe_{0.33}V_{0.15}$), the hydrogen pressure could reach almost 200 bar. They concluded that a three stage MHHC has better performance than just a two stage one. Wang et al. [11] developed a two-stage (40 bar to 700 bar) hydride compressor using $AB_2$ hydrides by comparing the thermodynamic behavior of different compositions of the $AB_2$ alloys Ti-Cr-Mn and Ti-Zr-Cr-Fe-V, for the first and second stage of the compressor respectively. Lototsky et al. [9] created a MH prototype compressor operating in the temperature range 20-150 ºC and providing compression of hydrogen from 10 to 200 bar with an average productivity up to 1000 l/h. The compressor has a two stage layout where the first part uses $AB_5$- and the second $AB_2$-type hydride-forming intermetallic compounds. Finally, other prototypes were described by

Lototsky et al. [3] that covered several ranges of temperatures and pressures to provide different system performances.

In our previous works [13 - 15], thermodynamic simulations and $AB_2$ alloys experimentation in single and three-stage MHHC were carried out to optimize the compression ratio and the amount of compressed hydrogen. In [13], models for real gas behavior and Pressure-Composition isotherms (P-c-I) curves were selected to be used in future simulations. In [14] the mechanical behavior of the $AB_2$ materials selected for a three stage MHHC during hydrogenation cycling was investigated defining the maximum filling value of a system based on these compounds. In [15], we selected three $AB_2$ materials from a database containing more than 200 compounds obtained from literature and by using a MATLAB code, we simulated the behavior of a three stage MHHC based on those compounds, obtaining relevant data for the compressor design, such us compression ratio, productivity, relationship between masses and volumes of the three stages. This can be considered the main innovation of our work as the use of realistic features in the simulation of the operation of the MHHC allow us to define and optimize confidently these parameters. In addition, it is worth mentioning that the MHHC has been designed to be coupled to a $H_2$-production method, photo-electrolysis, that use direct sun light to decompose water. In such method $H_2$ is produced at 1 bar and then our MHHC has a low equilibrium pressure hydride for the first stage.

In the present work, we have considered previous results to develop a three-stage MHHC prototype system using real operational parameters. First, a MHHC system is assembled in such a way that different materials for each stage and different heating and cooling methods can be easily investigated between vacuum and 250 bar. The volume of each stage has been calibrated and the thermal management is set up for the operational temperature ranges. Then, the operation of the MHHC system has been tested and a further analysis of the complete compression process is accomplished. The implementation of a test system like the one described in this work can aid to verify the optimal combination of different hydrides designed to achieve a proper final pressure from ambient conditions. In addition, the system allows to obtain relevant parameters of the compressor such as the compression ratio and the amount of $H_2$ gas moles compressed.

## 2. Methods

### 2.1 Design, assembly and set up of the MHHC system

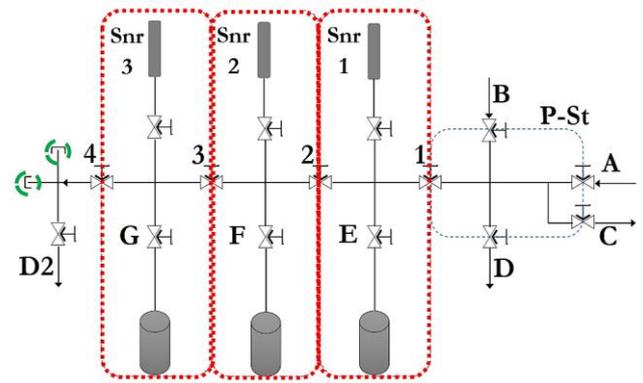

**Figure 1.** Design scheme of the MHHC system

The scheme of the MHHC system is displayed in Figure 1. From right to left, it shows first a pre-stage section (P-St, dash blue rectangle) where the control of all inlet (A: $H_2$, B: Ar) and outlet (D: atmosphere and C: vacuum) gases takes place. Then, the three compression stages (St) follow in series (dash red rectangles, where 1, 2 and 3 stand for the first, second and third stage, respectively). A security valve (4) is placed at the outlet of the third stage where the compressed gas can be stored on external volumes (dash green circles) or exhaust to the atmosphere (D2). The volumes of each stage (V, cross and pressure sensor volumes) are delimited by the inlet and outlet valves (e.g. 1 and 2), the connection valves (E, F, G) to the reactors with the alloys and a pressure sensor (Snr) at the top. The system is manually operated.

The whole system was assembled using Swagelok® connectors and MKS® pressure sensors. Leak tests were performed at 120 bar during more than 2 hours. All stage volumes (volume of the cross and sensor) were calibrated with errors under 5%, $V_1= 11.40 \pm 0.25$ cm$^3$, $V_2= 11.39 \pm 0.25$ cm$^3$, $V_3= 11.45 \pm 0.25$ cm$^3$. A larger error, 13%, occurred in the calibration of the pre-stage section but this volume can be isolated from the compression stages and thus it is not used during hydrogen compression experiments.

### 2.2 Thermal management system

The second most important part for the proper operation of the MHHC is the thermal system. It should fulfil three functions, *i)* to heat the hydrides during the desorption step, *ii)* to cool down the hydrides during the absorption step, and *iii)* maintain the temperature stable (high at desorption and low at absorption) to optimize the reaction kinetics. Considering these operational requests, the thermal management system entails a cooling and a heating system.

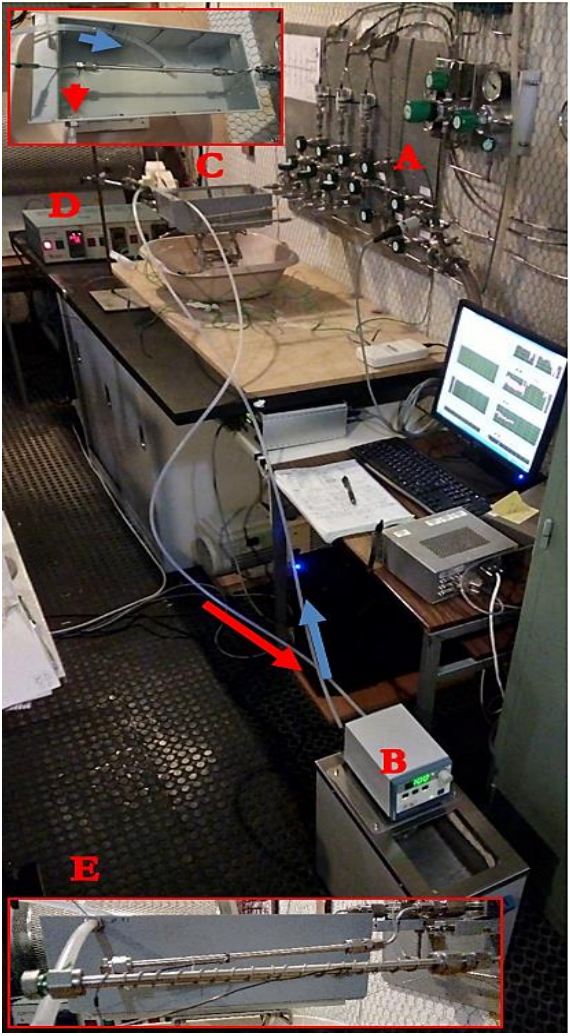

**Figure 2.** Picture of the MHHC test system, A. MHHC stages, B. water (23ºC) pump, C. submerged reactor in the thermal bath, D. PDI unit and E. reactor with a Thermocoax surrounded.

The cooling system (Figure 2. B and C) was employed to minimize temperature shifts in the stages during absorption steps. For this purpose, each stage is immersed in turn in a thermally controlled water bath (Figure 2. C). A high flux of deionized water driven by a mechanical pump with temperature regulation at 23°C (Figure 2. B), circulates across the water bath. This is represented by blue/red arrows at the inlet/outlet of the water bath, respectively. This guaranties that when temperature increases in the stage due to hydrogen absorption, the heat is fast dissipated by the water bath and stage temperature stabilized at 23ºC.

The heating management was controlled by a Thermocoax element (Figure 2. E) that is rolled around the reactor and heats it up to the target temperature of desorption $T_H = 120$ ºC. To achieve this temperature a Proportional-Derivative-Integral (PDI) control unit (Figure 2. D) regulates the electrical power with the aid of a thermocouple attached to the vessel, with minimum damping effects so that the time to attain these parameters is the lowest possible. Also, when a reduction in the set-point temperature is generated, the control unit adjusts the value promptly. To improve temperature regulation, the reactor was wrapped with aluminum foil.

## 2.3 Experimental measurements of the three stage MHHC system

For the final assembly of the MHHC, the materials selected were the same $AB_2$ alloys used in previous works [15] (a low B1, a medium B2 and a high B3 plateau pressure hydride, Table 1) but synthesized in larger quantities (≈100 g in total for each type) by induction melting. Morphological and microstructural characterisation of these alloys are described in ref [15]. The samples used in the present investigation were shortly milled, sieved with particle size between 150 μm and 1.2 mm and poured in the reactors (Figure 2. E). A horizontal tube design has been used to minimize the compactation/densification of metal hydride particles during hydrogen absorption/desorption cycles, which can be higher in a vertical design due to gravity [14], [16]. The amount of mass in each vessel was calculated to have a packing volume fraction of the hydrides of 63 vol.%, a value that is somewhat lower than that used in our previous simulations [14].

**Table 1.** Chemical composition of the three alloys used in the MHHC system

| | |
|---|---|
| B1 | $Ti_{0.85(1)}Zr_{0.15(1)}Mn_{1.41(1)}V_{0.31(1)}$ |
| B2 | $Ti_{0.80(1)}Zr_{0.20(1)}Mn_{1.28(1)}Cr_{0.60(1)}V_{0.20(1)}$ |
| B3 | $Ti_{0.90(1)}Zr_{0.10(1)}Mn_{1.48(1)}Cr_{0.40(1)}V_{0.20(1)}$ |

Additionally, using the same reactor volume for the three stages was an easier technical approach for the building of the compressor. Moreover, it simplifies the implementation of heating or cooling methods. However, as a drawback, several hydrogenation cycles are needed to fully charge the last stage of the MHHC. Therefore, the experiments were performed to analyze the behavior of several cycles in the final outcome. The calibrated volume of each reactor, i.e. just the volume where the sample is enclosed, ($V_{ri}$) and the mass of alloy ($M_i$) were: $V_{r1}= 4.5 \pm 0.1$ cm$^3$, $V_{r2}= 4.6 \pm 0.1$ cm$^3$, $V_{r3}= 4.6 \pm 0.1$ cm$^3$, $M_1= 11.010 \pm 0.001$ gr, $M_2= 11.070 \pm 0.001$ gr and $M_3= 10.700 \pm 0.001$ gr. All alloys were activated in three hydrogenation/dehydrogenation cycles before their normal use in the compressor

## 3. Results

### 3.1 Activation of alloys.

In Figure 3. the activation of all the alloys is displayed. In the left panel (Figure 3. a, c and e) raw data are provided, while the right panel (Figure 3. b, d and f), the hydrogenation state of these activation steps with respect to the equilibrium P-c-I data of the three alloys at $T_L = 23$ºC is shown.

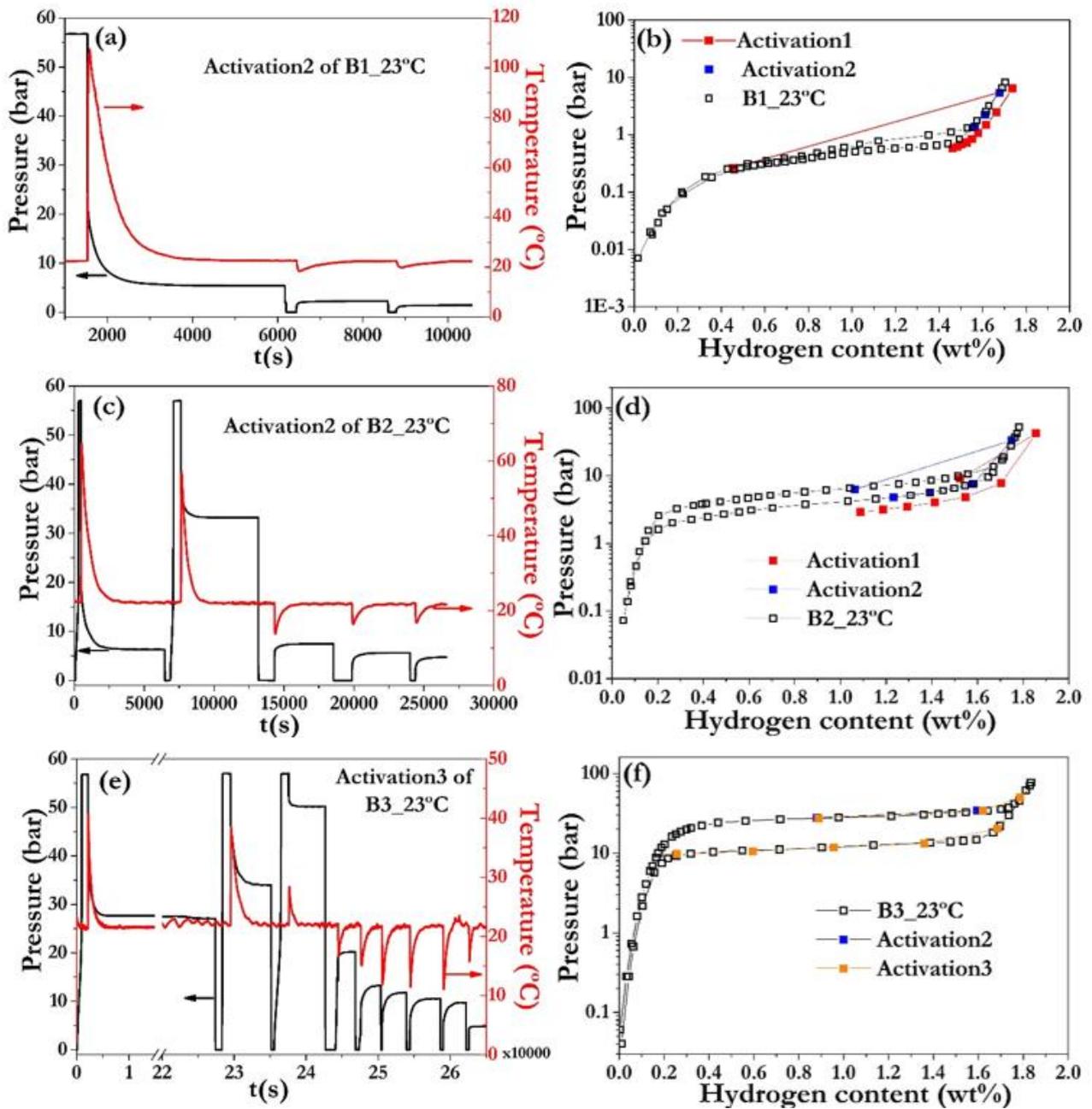

**Figure 3.** Activation of the three alloys (B1, B2, B3) at $T_L$ = 23ºC and their comparison to their respective P-c-I isotherms at the same temperature. Panels a, c, e) stand for specific activation data showing the time-evolution of pressure (black) and temperature (red). Panels b) d) and f) stand for pressure-composition activation data after reaching equilibrium (full squares) as compared to in their respective P-c-I isotherms (empty squares) for B1, B2 and B3 alloys, respectively. In e) the time is multiplied by 10000. No water-cooling system is used in these experiments.

The P-c-I curves displayed in Figure 3 match with those of the alloys studied in previous works for a simulation and design of a three-stage MHHC system [15]. An important feature observed in Figures 3 a b and c is the very large thermal drifts for all alloys due to the use of a large alloy mass. Temperature stabilization after absorption/desorption is slow. It should be noted that the experiments shown in Figure 3, were done without the water cooling system. Consequently, the stabilization time was very long, taking up to several hours (Figure 3(a)), which confirms that the (de-)hydrogenation kinetics are determined by heat transfer, *i.e.* the removal of heat in this hydrogenation stage. Therefore, a thermal management system is needed to improve the kinetics of the entire MHHC.

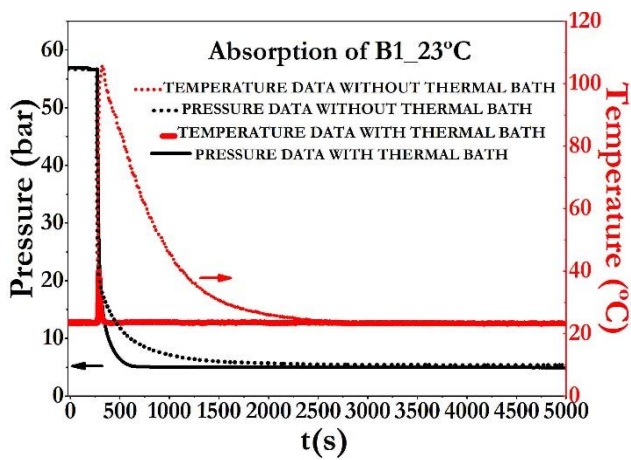

**Figure 4.** Effect of the thermal management system in the activation of the B1 sample (right red axis is the temperature).

Temperature drifts can be minimized by using the thermal bath. Thus, Figure 4 shows a clear reduction of the temperature peak for the B1 alloy by comparing the pressure and temperature when the activation of the alloy was done with (solid line) and without (dot line) a thermal bath. The circulating thermal bath reduces the temperature drift from ~84(1) to ~18(1)°C and the stabilization temperature time of the system from 75 min to approximately 1min. This variation in the temperature has a relevant effect on the absorption time, which for this case, reduces from 75 min without thermal bath to 9 min with this system. The heating system was tested at different temperatures to quantify its ability to perform the control of the temperature inside certain limits and over time. Fast heating (~50°C/min ) to the desorption temperature (120°C) can be achieved. This desorption temperature can be mantained constant inside a range of ±1°C over time.

With all the system tested and the alloys activated, the compression experiment using the three stages was performed.

### 3.2 Three stage MHHC system experiment

Figure 5 shows the pressure time-evolution in the three-stages (a) and the tracking of the absorption state of each alloy compared to their respective P-c-I absorption data (b) during the whole compression experiment. The experiment was performed in four hydrogenation cycles, from an initial pressure of 1.44 bar at 23°C in the first stage to approximately 120 bar at 120 °C in the last stage of the MHHC system. The need for more than one cycle to fully charge the last stage results from the fact that same alloy mass is used in each stage. According to previous works [15], these masses should have a ratio approximately of 1:2:3 for stages 1, 2 and 3, respectively, to compress hydrogen in one cycle.

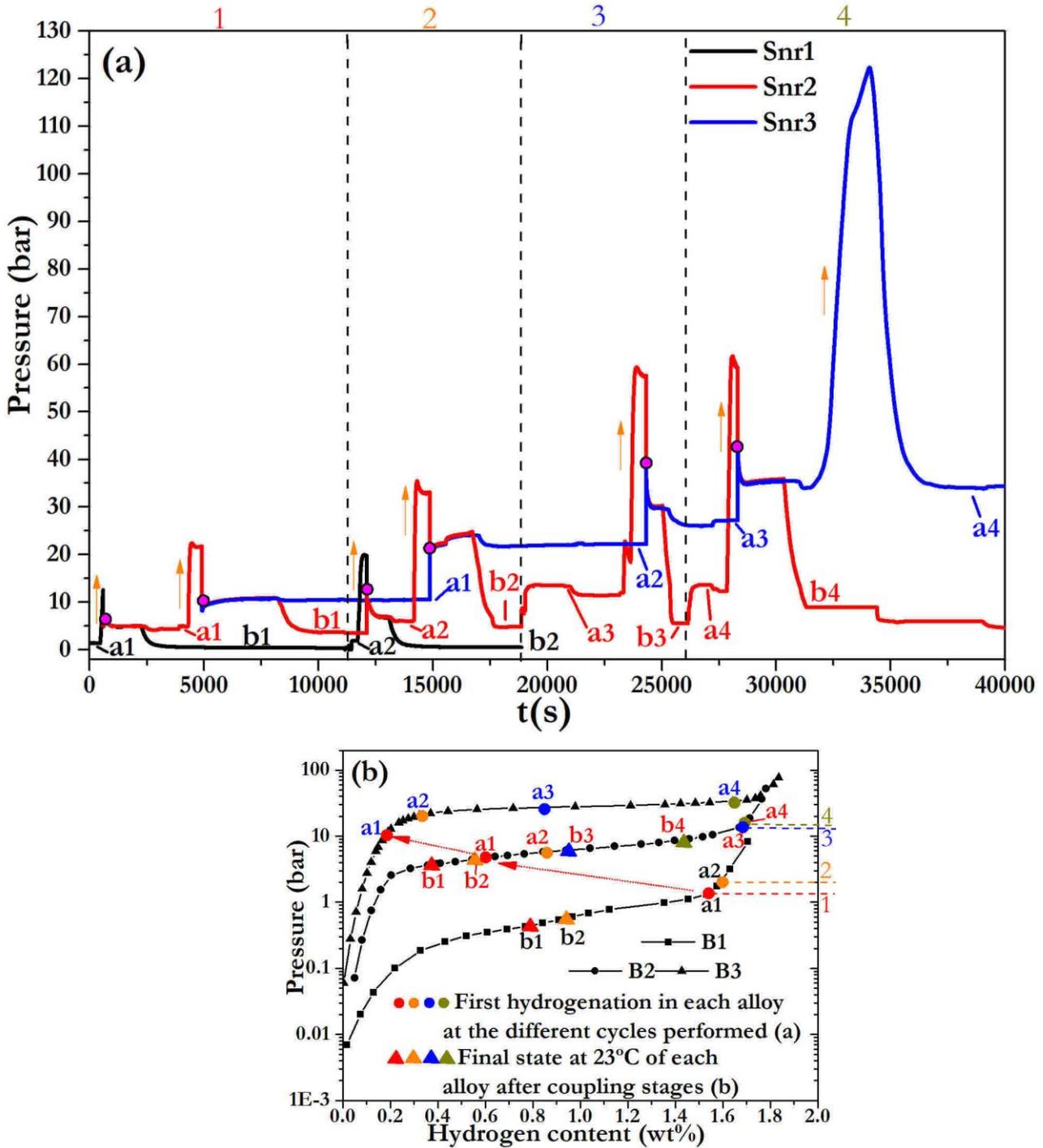

**Figure 5.** Experiment resume. (a) The operation and behavior of the three stage MHHC measurement system, (b) followed path in the absorption P-c-I's at $T_L = 23°C$. Each stage has a sensor (Snr) that will follow the evolution of the experiment, Snr1 (black line), Snr2 (red line), Snr3 (blue line). The color of each number and symbol stand for each cycle described in the experiment. Orange arrows in Fig.5a indicate when sensible heat is applied to the stage and magenta points indicate when connection of stages takes place.

All the data represented in Figure 5b correspond to an absorption state at $T_L$. Data points labelled as 'a' correspond to partially charged alloy in each stage, and, the 'b' label stand for a partially discharged alloy in all the stages (i.e. reabsorption at 23°C after the coupling of the stages and transfer of hydrogen, filled triangles). Both labels and their relative colours (black, red and blue) are related to the labels and colours in Figure 5a. In order to hydrogenate the second and third stages, several cycles were done from the initial hydrogenation in St1 to the hydrogenation in St3.

The process for the first cycle (red 1 on top of Figure 5a) is defined by red arrows connecting the path followed in Figure 5 (b). For this cycle, an inlet pressure $P_{in}$=1.44 bar in the first stage (black 'a' label) hydrogenates this alloy to the $\beta_{min}$ composition (Operational point (O.P.)) at $T_L$. Then, sensible heat is applied up to $T_H$ (orange arrows in Figure 5a), increasing the pressure in the first stage up to 12.5 bar. At this moment the second stage is at $T_L$. When temperature and pressure are stabilized in the first stage, the valve to the second stage is opened and a drop of pressure is evident due to the expansion of the gas and the simultaneous absorption of the second hydride (magenta point in Figure 5a). When the pressure stabilizes (red 'a' label) the valve separating the two stages is closed. Then, the heating system in stage 1 is switch-off, the temperature lowers to $T_L$ and the pressure drops to the re-absorption point labelled as black 'b' in Figure 5. Simultaneously, the O.P. for the second stage at $T_L$ is measured (red 'a' label in Figure 5). The same methodology is folowed between stages 2 and 3 achiving the O.P. for the 3 stage (blue 'a' label).

Cycles 2, 3, 4 (on top of Figure 5a), and labels in Figure 5: a2, b2, a3, b3 and a4, b4, correspond to additional cycles to achieve the final O.P. ("a4") in St3. This final point reached the target $P_{out}$ > 120 bar after heating at $T_H$, (section 4 on top of Figure 5(a). Indeed, a pressure value of 122 bar was obtained in St3.

## 4. Discussion

The design and set-up of the three stage MHHC measurement system fulfill all the requirements to perform the three stage MHHC system experiment, considering the calibration of all the volumes and also the thermal management system setting with low errors and thermal stabilization times

Considering that the inlet and outlet pressure are $P_{in}$ = 1.44 bar and $P_{out}$ = 122 bar, respectively, for a temperature window stablished between $T_L$ = 23 °C and $T_H$ = 120°C, the compression ratio CR of the MHHC measurement system is 84.7. This value is slightly higher than the one simulated in our previous works, CR= 75.9, with similar alloys and temperatures [15]. This result points out that the main objective of the work which was to be able to design and construct a MHHC based on $AB_2$-type alloys has been achieved and therefore validates the several approaches follow in this work and in previous ones.

The complete operation of the system, including the four cycles, took as long as 10 hours. This mainly results from the fact that the system is not automatic and its operation is not optimized in time as we choosed to atain full equilibrium conditions for the absorption – desorption in all the stages.

Some features of the thermal management of the MHHC are relevant to highlight. Figure 6 shows the evolution of pressure (a) and temperature (b) during coupling of st2 (red) and st3 (blue) in the 3$^{rd}$ and 4$^{th}$ cycle. Heating of st2 to 120ºC results in an almost instantaneous increase of pressure in st2. Also, the reaching of equilibrium during the coupling of stages (dot circles in 6 a and b) can be considered very fast ( < 700 s). This is mainly due to the reactor design in the form of a long tube of small thickness (< 2 mm) and diameter (< 8 mm). It results in low thermal resistance due to large heat exchange area and distance reduction between heat generation and sink [3]. This aids to reduce the thermal transient times due to a faster conduction of the heat from the thermal system (cooling and heating) to the inner parts of the compressor, i.e. hydride and $H_2$ gas [3], [17].

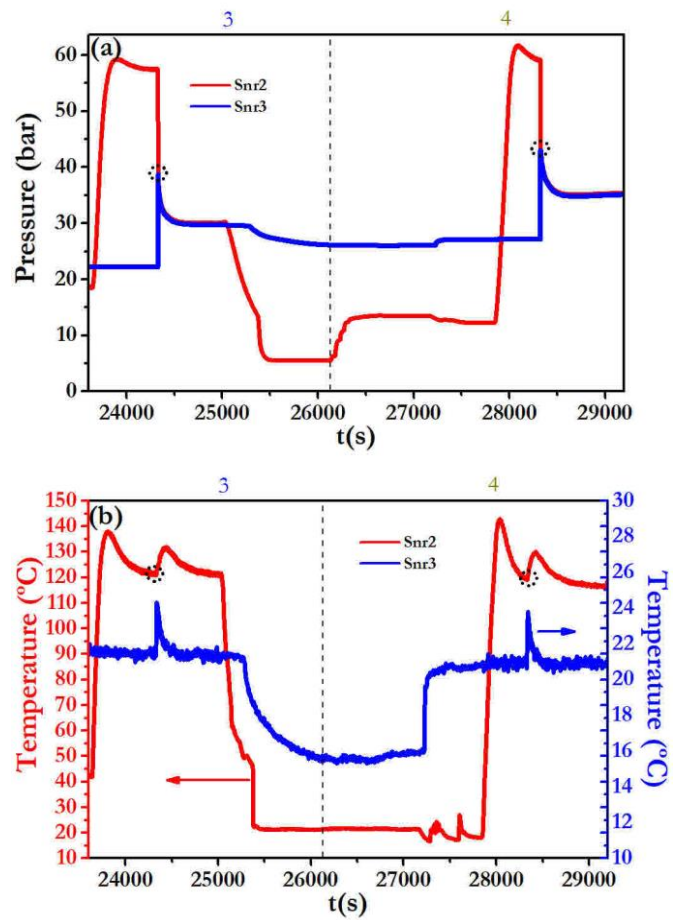

**Figure 6.** Coupling of St2 (red) and St3 (blue). Zoom between cycles 3 and 4 of Figure 5(a). (a) pressure data, (b)

temperature data. Dot circle is the conection point between stages.

In addition to this geometrical aspects, the heating/cooling system has also some relevance on the kinetics of the $H_2$ transfer between stages. As can be sen in Figure 6b, when the second and third stages are connected, shown by a dot circle in the figure, there is a respective decrease/increase of temperature due to the endothermic/exothermic nature of the desorption/absorption process that drastically counter act by the heating/cooling system implemented, helping to improve the kinetics of both processes.

It is also important to emphasize that the simplicity of the system (each reactor is a thin stainless steel tube) gives the opportunity to build a more complex MHHC system, i.e. a tube bundle, which can aid to increase the productivity of the system without complex changes in the thermal management system.

## 4. Conclusions

A three stage MHHC system has been successfully built and set-up in such a way that different materials for each stage and different heating and cooling methods can be easily investigated. The ranges of temperatures and pressures that can be measured go from RT to 200ºC and from vacuum to 250 bar, respectively.

The three stages have been prepared with the alloys selected and characterized through the previous works [10-12], where the geometry design has been chosen considering the results obtained through the whole study of the alloys and simulated system, specifically the porosity evolution of the hydrogenation cycles, the heat resistance of the system and hydrogenated packing fraction. Also, the simple geometry of the reactors aid to employ the thermal management system properly for an accurate control and measurement.

The MHHC system has been tested, giving several outcomes. A final experimental CR of 84.7 has been achieved at 120ºC in the system. Also, the simulation of the experimental MHHC behavior approached to these CR value with a similar alloy and temperature. The implementation of the $AB_2$-type alloys evidence that the time response during coupling of the stages is very fast, considering the intrinsic kinetics of the alloys and the thermal management employed. Also, the thermal control helps to abate the temperature excursions generated during the absorption and desorption processes. The simple geometry applied and validated in this study resembles a more complex and regular system like a tube bundle geometry design. Hence, an extrapolation of the current system to a bundle one can be easily implemented and could increase the productivity of the system.

**Acknowledgements**

Spanish MICINN financial support under contract (RTI2018-099794-B-I00) is acknowledged. The author A.R. Galvis E. would like to thank the "Departamento administrative de ciencia, tecnologia e innovacion - Colciencias-" for their financial and administrative support. Finally, to the ICMPE-CNRS crew in Thiais-Paris for their collaboration and aid in the development of the validating experiments of the alloys used.